\documentclass[12pt]{article}
\usepackage{amsmath}
\usepackage{amssymb}
\usepackage{epsfig}

\begin{document}
\def\be{\begin{equation}}
\def\ee{\end{equation}}
\def\ba{\begin{eqnarray}} 
\def\ea{\end{eqnarray}}
\def\nn{\nonumber}

\newcommand{\bbf}{\mathbf}
\newcommand{\rrm}{\mathrm}

\title{Site occupation constraints in mean-field approaches of quantum 
spin systems at finite temperature.\\} 
\author{Raoul Dillenschneider\footnote{E-mail address: rdillen@lpt1.u-strasbg.fr}
 and Jean Richert\footnote{E-mail address: richert@lpt1.u-strasbg.fr}\\ 
Laboratoire de Physique Th\'eorique, UMR 7085 CNRS/ULP,\\
Universit\'e Louis Pasteur, 67084 Strasbourg Cedex,\\ 
France} 
\date{\today}
\maketitle 
\begin{abstract}

We study the effect of site occupation on the description of quantum spin systems at finite temperature 
and mean-field level. We impose each lattice site to be occupied by a single electron. This is realized 
by means of a specific prescription. The outcome of the prescription is compared to the result obtained 
by means of a projection procedure which fixes the site occupation to one particle per site on an 
average. The comparison is performed for different representations of the Hamiltonian in Fock space leading 
to different types of mean-field solutions. The behaviour of order parameters is analyzed for each choice 
of the mean-field and constraint which fixes the occupation rate at each site. Sizable quantitative 
differences between the outcomes obtained with the two different constraints are observed.
\end{abstract} 
\maketitle
\\

\section{Introduction}

We consider ordered quantum spin systems at finite temperature in which each lattice site is occupied by 
one electron with a given spin. Such a configuration can be constructed by means of constraints imposed 
through a specific projection operation ~\cite{pop} which fixes the occupation in a strict sense.
The constraint can also be implemented on the average by means of a Lagrange multiplier procedure~\cite{aue}.
It is the aim of the present work to confront these two approaches in the framework of Heisenberg-type 
models.\\
 
The description of strongly interacting quantum spin systems at finite temperature generally goes
through a saddle point procedure which is a zeroth order approximation of the partition function.   
The so generated mean-field solution is aimed to provide a qualitatively realistic approximation 
of the exact solution.\\

However mean-field solutions are not unique. The implementation of a mean-field structure is for 
a large part subject to an educated guess which should rest on essential properties of the considered 
system, in particular its symmetries. This generates a major difficulty. A considerable amount of work 
on this point has 
been made and a huge litterature on the subject is available. In particular, systems which are 
described by Heisenberg-type models without frustration are seemingly well described by ferromagnetic
or antiferromagnetic (AF) N\'eel states at temperature $T = 0$ ~\cite{ber1,ber2}. It may however 
no longer be the case for many systems which are of low dimensionality ($d \leq 2$) and (or) frustrated 
~\cite{zha,mis,lec}. These systems show specific features. An extensive analysis and discussion in 
space dimension $d = 2$ has recently been presented by Wen ~\cite{wen}. The competition between AF 
and chiral spin state order has been the object of very recent investigations in the framework
of continuum quantum field approaches at $T = 0$ temperature, see ~\cite{tan,her}.\\ 

The reason for the specific behaviour of low dimensional systems may qualitatively be related to the
fact that low dimensionality induces strong quantum fluctuations, hence disorder which destroys the AF 
order. This motivates a transcription of the Hamiltonian in terms of composite operators which we call 
"diffusons" and "cooperons" below, with the hope that the actual symmetries different from those which 
are generated by AF order are better taken into account at the mean-field level ~\cite{aue}.\\ 

In the present work we aim to work out a rigorous vs. average treatment of the half-filling occupation
constraint on systems governed by Heisenberg-type Hamiltonians at the mean-field level, for different 
types of order.\\ 

The outline of the paper is the following. In section 2 we first recall the projection procedure
which leads to a rigorous site occupation. Section 3 is devoted to the confrontation of the magnetization
obtained through this procedure with the result obtained by means of an average projection
procedure in the framework of the mean-field approach characterized by a N\'eel state. The same 
confrontation is done in section 4 for the order parameter which characterizes the system when its 
Hamiltonian is written in terms of so called Abrikosov fermions~\cite{mis,aue} in $d = 2$ space 
dimensions. In section 5 we show that the rigorous projection ~\cite{pop} is no longer applicable when 
the Hamiltonian is written in terms of composite "cooperon" operators. Conclusions, further investigations 
and extensions are presented in section 6.\\

\section{Site occupation constraint for quantum spin systems at finite temperature T.}

Heisenberg quantum spin Hamiltonians of the type

\be
H = \frac {1}{2}\sum_{i,j} J_{ij} \vec S_{i} \vec S_{j}
\label{eq1}\ 
\ee
with $\{J_{ij}\} > 0$ can be projected onto Fock space by means of the transformation  

\ba\nonumber
S^{+}_{i} = a^{\dagger}_{i \uparrow} a_{i \downarrow}
\ea
\ba\nonumber 
S^{-}_{i} = a^{\dagger}_{i \downarrow} a_{i \uparrow}
\ea
\ba\nonumber 
S^{z}_{i} = \frac{1}{2} (a^{\dagger}_{\uparrow} a_{i \uparrow} - 
a^{\dagger}_{\downarrow} a_{i \downarrow})
\ea
where $\{a_{i \uparrow}, a_{i \downarrow}\}$ are anticommuting fermion operators.
\\  

This transformation is not bijective because the dimensionality of Fock space is larger than 
the dimensionality of the space in which the spin operators $\{\vec S_{i}\}$ are acting. Indeed, in 
Fock space, each site $i$ can be occupied by $0$, $1$ or $2$ fermions corresponding to the states 
$|0,0>,|1,0>,|0,1>,|1,1>$ where $|0,0>$ is the particle vacuum, $|1,0> = |+ 1/2>$ ,$|0,1> = |- 1/2>$ 
and $|1,1> = |+ 1/2, - 1/2>$ in terms of spin $1/2$ projections. Since one wants to 
keep states with one fermion per site only states $|0,0>$ and $|1,1>$ have to be 
eliminated. This can be performed on the partition function for a system at inverse 
temperature $\beta$

\ba\nonumber
Z = Tr e^{-\beta H}
\ea
where the trace is taken over the whole Hilbert space by adding a projection term

\be
Z = Tr e^{-\beta (H - \mu N)}
\label{eq2}\  
\ee
where $N$ is the particle number operator and $\mu = i \pi/ 2 \beta$ an imaginary chemical 
potential ~\cite{pop}.
Indeed, the presence of the states $|0,0>_{i}$ and $|1,1>_{i}$ on site $i$ leads in  $Z$ to  
phase contributions which eliminate each other 
\be
e^{i*0} + e^{i* \pi} = 0 
\label{eq3}\ 
\ee
and hence the contributions of these spurious states are cancelled as a whole.\\

The common alternative approximate projection procedure would be to introduce a chemical potential
in terms of $real$ Lagrange multipliers $\{\lambda_{i}\}$ and to fix this quantity by means of a saddle 
point procedure with respect to the $\{\lambda_{i}\}$ s

\be
Z = Tr e^{-\beta H} {\prod_{i} \int d\lambda_{i}  e^{\lambda_{i}(n_{i}-1)}}
\label{eq4}\  
\ee
where $n_{i}$ is the particle number operator on site $i$.\\ 

In the following we compare the outcome of these two projection procedures in the framework of different
mean-field approximations of systems described by Heisenberg Hamiltonians.\\

\section{Antiferromagnetic mean-field ansatz}

\subsection{Exact occupation procedure}

Starting with the Hamiltonian defined in Eq.~(\ref{eq1}) the partition function $Z$
can be written in the form
\be
Z = \int \prod_{i, \sigma}{\mathcal{D}} (\{\xi^{*}_{i, \sigma},\xi_{i,\sigma}\})
e^{- \, A(\{\xi^{*}_{i,\sigma},\xi_{i,\sigma}\})} 
\label{eq5}\
\ee 
where the $\{\xi^{*}_{i,\sigma},\xi_{i,\sigma}\}$ are Grassmann variables corresponding to the 
operators $\{a^{\dagger}_{i \sigma}, a_{i \sigma}\}$ defined above. They depend on the imaginary 
time $\tau$ in the interval $[0,\beta]$. The action $A$ is given by
\be
A(\{\xi^{*}_{i,\sigma},\xi_{i,\sigma}\}) = \int_{0}^{\beta} d\tau \sum_{i,\sigma} (\xi^{*}_{i,\sigma} 
(\tau) \partial_{\tau} \xi_{i,\sigma} (\tau) + {\cal H}(\{\xi^{*}_{i,\sigma} (\tau),\xi_{i,\sigma} 
(\tau)\}))
\label{eq6}\ 
\ee
where 
\be
{\cal H} (\tau) = H(\tau) - \mu N(\tau)
\label{eq7}\ 
\ee
and $N(\tau)$ is the particle number operator. A Hubbard - Stratonovich (HS) transformation 
which generates the vector fields $\{\vec {\varphi}_i\}$ leads to the partition function $Z$ 
which can be written in the form
\be
Z = \int \prod_{i, \sigma}{\mathcal{D}} (\{\xi^{*}_{i, \sigma},\xi_{i,\sigma},
\vec {\varphi}_i\})
e^{- \int_{0}^{\beta} d\tau  [\sum_{i,\sigma} \xi^{*}_{i,\sigma} 
(\tau) \partial_{\tau} \xi_{i,\sigma} (\tau) + \tilde H(\{\xi^{*}_{i,\sigma},
\xi_{i,\sigma}, \vec {\varphi}_i\})]} 
\label{eq8}\ 
\ee 
In Eq.~(\ref{eq8}) the expression of $Z$ is quadratic in the Grassmann variables 
$\{\xi^{*}_{i,\sigma},\xi_{i,\sigma}\}$ over which the expression can be integrated. The remaining 
expression depends on the fields $\{\vec {\varphi}_i(\tau)\}$. A saddle point procedure decomposes 
$\vec{\varphi}_i(\tau)$ into a mean-field contribution and a fluctuating term
\be
\vec {\varphi}_i(\tau) = \vec{\varphi}_{j}^{(mf)}  + \delta \vec {\varphi}_i(\tau)
\label{eq9}\ 
\ee 
where $\vec{\varphi}_{i}^{(mf)}$ are the constant solutions of the self-consistent equation
\be 
\vec{\varphi}_{i}^{(mf)}  = \frac{1}{2} \sum_{j} J_{ij}
\frac{\vec{\varphi}_{j}^{(mf)}} {\| \vec{\varphi}_{j}^{(mf)}\|} th \,
\left( \frac{\beta \|\vec{\varphi}_{j}^{(mf)} \|}{2} \right) 
\label{eq10}\ 
\ee
\\
The partition function takes the form
\be 
Z = Z_{mf} \int {\mathcal{D}} (\{\delta \vec {\varphi}_i\})e^{- A(\{\delta \vec {\varphi}_i\}} 
\label{eq11}\ 
\ee
where the first term on the r.h.s. corresponds to the mean-field contribution and the second term 
describes the contributions of the quantum fluctuations.\\

Considering the mean-field contribution the partition function $Z_{mf}$ can be put in the form
\ba\nonumber 
Z_{mf} = Tr e^{- \beta H_{mf}}
\ea
which reads  
\be
Z_{mf} = i^{-{\cal N}} e^{\frac{1}{2} \beta \underset{ij}{\sum} (J^{-1})_{ij} \vec{\varphi}_{i}^{(mf)} 
\vec{\varphi}_{j}^{(mf)}}  \prod_{i} (1 + e^{-\beta E^{+}_{i}})(1 + e^{-\beta E^{-}_{i}})
\label{eq12}\ 
\ee
where ${\cal N}$ is the number of sites. The energies $E^{+}_{i}$ and $E^{-}_{i}$ are obtained through 
a diagonalization of the Hamiltonian by means of a Bogolioubov transformation in Fock space. 
Explicitly
\ba\nonumber
E^{+}_{i} = \mu + \frac{\| \vec{\varphi}_{j}^{(mf)}\|}{2}
\ea
\ba\nonumber
E^{-}_{i} = \mu - \frac{\| \vec{\varphi}_{j}^{(mf)}\|}{2}
\ea
Working out the expression of $Z_{mf}$ leads to 
\be 
Z_{mf} =  e^{\frac{1}{2} \beta \underset{ij}{\sum} (J^{-1})_{ij} \vec{\varphi}_{i}^{(mf)} 
\vec{\varphi}_{j}^{(mf)}}  \prod_{i} ch \left(\frac {\beta \|\vec{\varphi}_{i}^{(mf)}\|}
{2}\right)               
\label{eq13}\ 
\ee
\\
and the free energy is given by the expression 
\be 
{\cal F}_{mf} = - \frac{1}{2} \underset{ij}{\sum} (J^{-1})_{ij} \vec{\varphi}_{i}^{(mf)} 
\vec{\varphi}_{j}^{(mf)} - \frac{1}{\beta} \sum_{i} \log 2 
ch \left(\frac{\beta \|\vec{\varphi}_{i}^{(mf)}\|}{2}\right) 
\label{eq14}\ 
\ee
 
In order to obtain the expression of the local magnetization $\{\vec {m}_{i}\}$ one should add 
the term $\sum_{i}\vec{B}_{i} \vec{S}_{i}$ to the Hamiltonian $H$ in Eq.~(\ref{eq1}). Going through 
the same steps as above the local magnetization given by
\be
\vec m_{i}^{(mf)} = - \frac{\partial {\cal F}_{mf}} {\partial \vec{B}_{i}}
 {\Big\vert}_{\vec{B}_{i}=\vec{0}} 
\label{eq15}\
\ee
is related to the $\{\vec{\varphi}_i^{(mf)}\}$'s by 

\ba\nonumber
\vec{\bar{m}}_{i} 
= \frac{1}{2} \frac{\vec{\bar{\varphi}}_{i}} {\bar{\varphi}_{i} }
th \left[ \frac{\beta\bar{\varphi}_{i}} {2} \right]
\ea

which leads to the self-consistent equation for the $\{\vec {m}_{i}^{(mf)}\}$'s
 
\be
\vec m_{i}^{(mf)} = \frac{2}{\beta} \sum_{j} (J^{-1})_{ij} (th^{-1} (2 m_{j})) \frac{\vec {m}_{j}^{(mf)}}
{m_{j}^{(mf)}}
\label{eq16}\   
\ee 
If the local fields $\{\vec{B}_{i}\}$ are oriented along a fixed direction $\vec {e}_{z}$,  
$\vec {m}_{i}^{(mf)} =  m_{i}^{(mf)} \vec {e}_{z}$, the magnetizations are the solutions
of the self-consistent equations 
\be
m_{i}^{(mf)} = \frac{1}{2} th \left(\frac{\beta \sum_{j} J_{ij} m_{j}^{(mf)}}{2}\right)
\label{eq17}\   
\ee 
\\

\subsection{Lagrange multiplier approximation}

Similarly to the preceding case one can introduce the one-particle site occupation by 
means of a Lagrange procedure added on the expression of the Hamiltonian  

\ba\nonumber
H =  \frac{1}{2} \sum_{i,j} J_{ij} \vec S_{i} \vec S_{j} + \lambda \sum_{i} (n_{i} - 1)
\ea
where $\lambda$ is a variational parameter and $\{{n_i}\}$ are particle number operators.\\

Following the same lines as above with the help of a HS transformation and staying in 
the ordinary space representation the mean-field partition function $Z_{mf}^{\lambda}$
can be worked out and reads
\be
Z_{mf}^{\lambda} = e^{-\frac{1}{2} \beta \underset{ij}{\sum} (J^{-1})_{ij} \vec{\varphi}_{i}^{(mf)} 
\vec{\varphi}_{j}^{(mf)} - {\cal N} \lambda}  
\prod_{i} (1 + e^{-\beta E^{+}_{i \lambda}})(1 + e^{-\beta E^{-}_{i \lambda}})
\label{eq18}\ 
\ee
\\
with

\ba\nonumber
E^{+}_{i \lambda }  = \lambda + \frac{\| \vec{\varphi}_{j}^{(mf)}\|}{2}
\ea
\ba\nonumber
E^{-}_{i \lambda }  = \lambda - \frac{\| \vec{\varphi}_{j}^{(mf)}\|}{2}
\ea

The parameter $\lambda$ is fixed through a minimization of the corresponding free energy 
${\cal F}_{mf}^{(\lambda)}$ with respect to this multiplier. The minimization shows that 
the extremum solution is obtained for $\lambda = 0$ and
\be
{\cal F}_{mf}^{(\lambda)} = - \frac {1}{2} \underset{ij}{\sum} (J^{-1})_{ij} \vec{\varphi}_{i}^{(mf)} 
\vec{\varphi}_{j}^{(mf)} - \frac{2}{\beta} \sum_{i} 
\ln 2 ch \left(\frac{\beta \|\vec{\varphi}_{i}^{(mf)}\|}{4}\right) 
\label{eq19}\
\ee
\\
which is different from the expression of Eq.~(\ref{eq14}).\\

The magnetization can be obtained in the same way as done above. One obtains   
\be
m_{i, \lambda = 0}^{(mf)} = \frac{1}{2} th \left(\frac{\beta \sum_{j} J_{ij} 
m_{j, \lambda = 0}^{(mf)}}{4}\right)
\label{eq20}\   
\ee 
which is again different from the expression obtained in the case of a rigorous projection, see
Eq.~(\ref{eq17}).\\

The uniform solutions $m_{i}^{(mf)} = (-1)^i m^{(mf)}$ 
and $m_{i \lambda = 0}^{(mf)} =(-1)^i m_{\lambda = 0}^{(mf)}$ for $\{J_{ij}\} = J$
have been calculated by solving the selfconsistent equations (17) and (20).
 
\begin{figure}
\centering
\epsfig{file=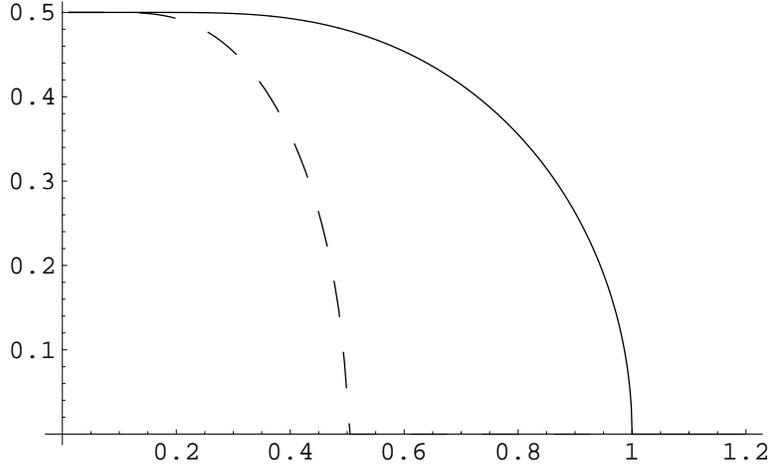}
\caption{Magnetization vs. reduced temperature $t= T / |J|$.
Full line: exact site occupation. Dashed line: average site occupation.}
\label{fig1}
\end{figure}

The results are shown in Fig.(\ref{fig1}). It is seen that the treatment of the site occupation affects 
sizably the quantitive behaviour of observables. In particular it shifts the location of the critical
temperature $T_c$ by a factor 2. Such a strong effect has already been observed on the behaviour of 
the specific heat, see refs.~\cite{aza1,aza2}.\\

\section{Spin state mean-field ansatz in 2d}

In 2$d$ space the Heisenberg Hamiltonian given by Eq.(1) can be written in terms of composite 
non-local operators $\{{\cal D}_{ij}\}$ ("diffusons") ~\cite{aue} defined as
\ba\nonumber
{\cal D}_{ij} = a_{i \uparrow}^{\dagger} a_{j \uparrow} + 
a_{i \downarrow}^{\dagger} a_{j \downarrow}
\ea
\\
Then the initial Hamiltonian Eq.~(\ref{eq1}) with $J_{ij} = J$ takes the form
\be
H = - J \sum_{<ij>} (\frac{1}{2} {\cal D}^{\dagger}_{ij} {\cal D}_{ij} - 
\frac{n_i}{2} + \frac{n_i  n_j}{4})  
\label{eq21}\ 
\ee
where $i$ and $j$  are nearest neighbour sites. The physical justification for such a transcription
relies on the fact that systems with low space dimensionality at finite temperature show physical 
symmetry properties which are closer to a spin liquid than an ordered N\'eel state as already 
mentioned in the introduction.\\  

The number operator products $\{n_i  n_j\}$ in Eq.~(\ref{eq21}) are quartic in terms 
of creation and annihilation operators in Fock space. In principle its existence necessitates
the introduction of a further mean field. One can however show that its elimination has no
influence on the results obtained from the partition function. As a consequence we leave it out
from the beginning as well as the contribution corresponding to the $\{n_i\}$ term.\\

\subsection{Exact occupation procedure}

Starting with the Hamiltonian
\be
H = - \frac{J}{2} \sum_{<ij>}   {\cal D}^{\dagger}_{ij} {\cal D}_{ij} - \mu N 
\label{eq22}\ 
\ee
one performs an HS transformation on the corresponding functional integral partition function in which 
the action contains the occupation number operator as written out in Eq.~(\ref{eq7}). The Hamiltonian   
takes the form
\be
H = \frac{2}{|J|}\underset{<ij>}{\sum} \bar{\Delta}_{ij} \Delta_{ij} 
+\underset{<ij>}{\sum} \left[ \bar{\Delta}_{ij} {\cal D}_{ij} +
\Delta_{ij} {\cal D}^{\dagger}_{ij} \right]  - \mu N
\label{eq23}\ 
\ee
\\
The fields $\Delta_{ij}$ and their conjugates $\bar \Delta_{ij}$ can be decomposed into a mean-field 
contribution and a fluctuation term
\be 
\Delta_{ij} = \Delta_{ij}^{mf} + \delta \Delta_{ij}
\label{eq24}\ 
\ee
\\

The field $\Delta_{ij}^{mf}$ can be chosen as a complex quantity 
$\Delta_{ij}^{mf}=|\Delta_{ij}^{mf}|e^{i\phi_{ij}^{mf}}$. Consider a square plaquette 
$(\vec i, \vec i + \vec {e}_{x},\vec i + \vec {e}_{y},\vec i + \vec {e}_{x} + \vec {e}_{y})$ 
where $\vec {e}_{x}$ and $\vec {e}_{y}$ are the unit vectors along the directions $\vec {Ox}$ 
and $\vec {Oy}$ starting from site $\vec i$ on the lattice. On this plaquette   
\ba\nonumber
\phi = \prod_{(ij) \in pl}\phi_{ij}^{mf}
\ea
is taken to be constant. If the gauge phase $\phi_{ij}^{mf}$ fluctuates in such a way that $\phi$ 
keeps constant the average of $\Delta_{ij}^{mf}$ will be equal to zero in agreement with 
Elitzur's theorem ~\cite{el1}. In order to guarantee $SU(2)$ invariance of the mean-field Hamiltonian
along the Wilson loop on the plaquette we follow ~\cite{wen,aff1,aff2,awe,lee} and introduce

\be
\phi_{ij}=
\begin{cases}
e^{i.\frac{\pi}{4}(-1)^i}, \text{if } \vec{r}_j=\vec{r}_i+\vec{e}_x \\
e^{-i.\frac{\pi}{4}(-1)^i}, \text{if } \vec{r}_j=\vec{r}_i+\vec{e}_y \\
\end{cases}
\ee

Then the total flux through the plaquette

\ba\nonumber
\phi = \pi 
\ea

In the mean-field approximation the corresponding partition function reads 
\be 
Z_{mf} = e^{- \beta (H_{mf} - \mu N)}
\label{eq27}\ 
\ee
where 
\be
H_{mf} = \frac{2}{|J|}\underset{<ij>}{\sum} 
\bar \Delta_{ij}^{mf}.\Delta_{ij}^{mf} + \underset{<ij>}{\sum} \left[
\bar \Delta_{ij}^{mf} {\cal D}_{ij} + 
\Delta_{ij}^{mf} {\cal D}^{\dagger}_{ij} \right]  - \mu N
\label{eq28}\ 
\ee
\\

Performing a Bogolioubov transformation which diagonalizes the remaining expression in Fourier space 
leads to 

\be
H_{mf} = \frac{{\cal N} z \Delta^{2}} {|J|} + \sum_{\vec k, \sigma} [E^{+}_{\vec k, \sigma}  
\beta^{\dagger}_{(+) \vec k, \sigma} \beta_{(+) \vec k, \sigma}
+ E^{-}_{\vec k, \sigma} \beta^{\dagger}_{(-) \vec k, \sigma} \beta_{(-) \vec k, \sigma}]
\label{eq29}\ 
\ee
where $z=4$ is the coordination and ${\cal N}$ the number of sites.
The momenta $\{\vec k\}$ act in the first half Brillouin zone (spin Brillouin zone) and the 
operators $\{\beta^{\dagger}, \beta\}$  are the transformed of the $\{a^{\dagger}, a\}$ 's through 
the rotation which leads to the diagonal expression Eq.~(\ref{eq29}). The eigenenergies
$E^{+}_{\vec k, \sigma}$ and $E^{-}_{\vec k, \sigma}$ are given by
 
\be
E^{+}_{\vec k, \sigma} = -\mu + 2 \Delta [cos^{2}(k_{x}) + cos^{2}(k_{y})]^{1/2}  
\label{eq30}\    
\ee
and similarly
\be
E^{-}_{\vec k, \sigma} = -\mu - 2 \Delta [cos^{2}(k_{x}) + cos^{2}(k_{y})]^{1/2}
\label{eq31}\ 
\ee
\\
The partition function $Z_{mf}$ has the same structure as the corresponding partition function in
Eq.~(\ref{eq18}) and the free energy is given by 
\be
{\cal F}_{mf} = \frac{{\cal N} z \Delta^{2}} {|J|} - \frac{1} {\beta} \sum_{\vec k, \sigma} 
\ln 2 \left(ch \beta \Delta \epsilon_{\vec k}\right)
\label{eq32}\
\ee
with 
\be
\epsilon_{\vec k} = 2 [cos^{2}(k_{x}) + cos^{2}(k_{y})  ]^{1/2}
\label{eq33}\
\ee
\\
Finally the variation of ${\cal F}_{mf}$ with respect to $\Delta$ leads to the self-consistent mean-field 
equation  
\be
\tilde \Delta = \frac{1} {2 {\cal N}} \sum_{\vec k, \sigma} \epsilon_{\vec k} 
th \left(\frac{\beta|J|\epsilon_{\vec k} \tilde \Delta} {z}\right)
\label{eq34}\
\ee
with $\tilde \Delta = z \Delta/|J|$.\\

\subsection{Lagrange multiplier approximation}

Similarly to Eq.~(\ref{eq23}) one may introduce a Lagrange constraint and write
\be
H^{(\lambda)} = \frac{2}{|J|} \underset{<ij>}{\sum} \bar \Delta_{ij} 
\Delta_{ij} + \underset{<ij>}{\sum} \left( \bar \Delta_{ij} {\cal D}_{ij} 
+ \Delta_{ij} {\cal D}^{\dagger}_{ij} \right) + \sum_{i}\lambda_i  
(n_{i} - 1)
\label{eq35}\ 
\ee
\\
Then for $\lambda_{i} = \lambda$
\be
H_{mf}^{(\lambda)} = \frac{{\cal N} z \Delta^{2}} {|J|} + \sum_{\vec k, \sigma} 
\left(E^{+\lambda}_{\vec k, \sigma}  \beta^{\dagger}_{(+) \vec k, \sigma} \beta_{(+) \vec k, \sigma}
+ E^{-\lambda}_{\vec k, \sigma} \beta^{\dagger}_{(-) \vec k, \sigma} 
\beta_{(-) \vec k, \sigma}\right) 
\label{eq36}\ 
\ee
with the eigenenergies
\be
E^{+\lambda}_{\vec k, \sigma} = \lambda + 2 \Delta [cos^{2}(k_{x})  +  cos^{2}(k_{y})]^{1/2}  
\label{eq37}\    
\ee
and similarly
\be
E^{-\lambda}_{\vec k, \sigma} = \lambda - 2 \Delta [cos^{2}(k_{x})  +  cos^{2}(k_{y})]^{1/2}
\label{eq38}\ 
\ee
\\

The expression of the free energy is now given by 
\be
{\cal F}_{mf}^ {(\lambda)} = -{\cal N}\lambda + \frac{{\cal N} z \Delta^{2}} {|J|} -
\frac{1} {\beta} \sum_{\vec k, \sigma} \ln [1 + e^ {- \beta E^{+\lambda}_{\vec k, \sigma}}]
[1 + e^ {- \beta E^{-\lambda}_{\vec k, \sigma}}]               
\label{eq39}\ 
\ee
\\

The minimization of this expression in terms of $\lambda$ delivers the solution $\lambda = 0$ 
and 

\be
{\cal F}_{mf}^ {(\lambda)} = -{\cal N}\lambda + \frac{{\cal N} z \Delta^{2}} {|J|} -
\frac{1} {\beta} \sum_{\vec k, \sigma} 2 \ln \left( 2 ch \frac {\beta \Delta \epsilon_{\vec k}}
{2}\right)                
\label{eq40}\ 
\ee
\\

The variation of ${\cal F}_{mf}^ {\lambda}$ with respect to $\Delta$ leads to the self-consistent 
mean-field equation 
equation  
\be
\tilde \Delta ^ {(\lambda)}= \frac{1} {{\cal N}} \sum_{\vec k, \sigma} \epsilon_{\vec k} 
th \left( \frac {\beta|J|\epsilon_{\vec k} \tilde \Delta^{(\lambda)}} {2 z}\right)
\label{eq41}\
\ee
with $\tilde \Delta ^ {(\lambda)} = z \Delta/|J|$.
\\
Expressions in Eq.~(\ref{eq40}) and  Eq.~(\ref{eq41}) should be compared to the expressions 
obtained in Eq.~(\ref{eq32}) and Eq.~(\ref{eq34}). Fig. 2 shows the behaviour of $\tilde \Delta $
for the two different treatments of site occupation on the lattice.
\\ 

\begin{figure}
\centering
\epsfig{file=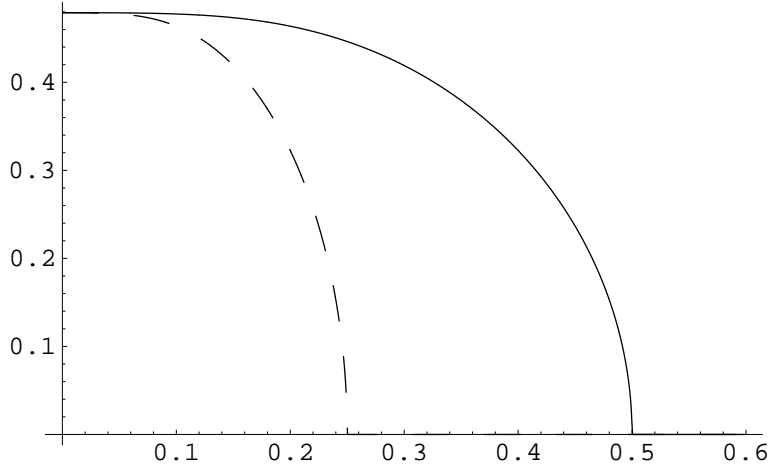}
\caption{$\tilde{\Delta}$ vs. reduced temperature $\tilde{t}=zT/|J|$.
Full line: exact site occupation. Dashed line: average site ocupation.}
\label{fig2}
\end{figure}

\subsection{Magnetization}

We calculate the magnetization in the case of a rigorous one-particle site 
occupation. Starting with the Hamiltonian $H$ given by Eq.~(\ref{eq22}) and adding the
staggered interaction $h \sum_{i} (-1)^{i}S^{z}_{i}$ of the spins with an homogeneous external field $h$
one obtains the mean field expression 
\be
H_{mf} = \frac{{\cal N} z \Delta^{2}} {|J|} + \sum_{\vec k, \sigma} [E^{+}_{h,\vec k, \sigma}  
\beta^{\dagger}_{(+) \vec k, \sigma} \beta_{(+) \vec k, \sigma}
+ E^{-}_{h, \vec k, \sigma} \beta^{\dagger}_{(-) \vec k, \sigma} \beta_{(-) \vec k, \sigma}]
\label{eq42}\ 
\ee
where
\be
E^{+}_{h, \vec k, \sigma} = -\mu + \tilde E_{h, \vec k, \sigma}   
\label{eq43}\
\ee
and similarly
\be 
E^{-}_{h, \vec k, \sigma} = -\mu - \tilde E_{h, \vec k, \sigma}  
\label{eq44}\
\ee
\\
with 
\ba\nonumber
\tilde E_{h, \vec k, \sigma} = \left(\frac {h^{2}}{4} + \Delta^{2} \epsilon^{2}_{\vec k}\right)^{1/2}
\ea
\\
The free energy can now be written
\be
{\cal F}_{h, mf} = \frac {{\cal N} z \Delta^{2}} {|J|} - \frac{1} {\beta} \sum_{\vec k, \sigma} 
\ln  \left(2 ch \beta \tilde E_{h, \vec k, \sigma}\right) 
\label{eq45}\
\ee
and the magnetization 
\ba\nonumber
m^{(mf)} = - \frac{\partial {\cal F}_{mf}} {\partial h}
{\Big\vert}_{h = 0}  
\ea
reads
\be
m^{(mf)} = \frac{\partial \Delta}{\partial h}{\Big\vert}_{h = 0} \left(\frac{2{\cal N} z \Delta}
{|J|} - \sum_{\vec k, \sigma} \epsilon_{\vec k} th (\beta \epsilon_{\vec k})\right) 
\label{eq46}\ 
\ee
The sum over $\vec k$ runs over half the Brillouin zone. Going back to Eq.~(\ref{eq34}) one sees that 
$m^{(mf)}$ is equal to zero.\\
The same result holds in the case of the average Lagrange multiplier procedure.
\\

\section{Cooperon mean-field ansatz}

Starting from  the Hamiltonian 
\be
H =   - |J| \sum_{<i,j>} \vec S_{i} \vec S_{j} 
\label{eq47}\
\ee
one can introduce a further set of non-local composite operators ("cooperons")
\be
{\cal C}_{ij} =   a_{i \uparrow} a_{j \downarrow} - a_{i \downarrow} a_{j \uparrow}
\label{eq48}\
\ee
This leads to the expression 
\be
H  = \frac {|J|}{2} \sum_{<i,j>} \left({\cal C}_{ij}^{\dagger} {\cal C}_{ij}  + 
\frac {n_{i} n_{j}} {2}\right) 
\label{eq49}\
\ee
where $n_{i} = \sum_{\sigma} n_{i \sigma}$.\\

\subsection{Exact occupation procedure}

As in the preceding cases it is possible to implement a HS procedure on the $\{{\cal C}_{ij}\}$ 
and $\{n_{i}\}$ in such a way that the expresssion of the corresponding partition function
gets quadratic in the fields $\{a_{i \uparrow}, a_{i \downarrow}\}$. The corresponding HS fields are 
$\{\Gamma_{ij}\}$, $\{\nu_{i}\}$ and 
\be
H - \mu N  = \frac {2}{|J|} \sum_{<i,j>} \bar\Gamma_{ij} \Gamma_{ij}
+ \sum_{<i,j>} \left(\bar\Gamma_{ij} {\cal C}_{ij} + \Gamma_{ij} {\cal C}_{ij}^{\dagger}
+ \frac {2}{|J|} \nu_{i}\nu_{j}\right) + \sum_{i} \nu_{i} n_{i}   
\label{eq50}\
\ee
\\
Introducing the homogeneous fields $\{\Gamma = \Gamma_{ij}\}$ ,$\{\nu = \nu_{i}\}$
and integrating over the cooperon fields $\{{\cal C}_{ij}\}$ one gets in Fourier space
\ba\label{eq51}
H &=& \frac{{\cal N} z |\Gamma|^{2}} {|J|} + \frac {2{\cal N}|\nu|^{2}} {z |J|}
+ \sum_{\vec k, \sigma} (\nu - \mu) a^{\dagger}_{\vec k, \sigma}
a_{\vec k, \sigma} 
\\ \nn
&  & + \sum_{\vec k, \sigma} \frac{\sigma z}{2} \gamma_{\vec k} (\bar\Gamma
a_{\vec k, \sigma} a_{-\vec k, -\sigma} - \Gamma
a^{\dagger}_{\vec k, \sigma} a^{\dagger}_{-\vec  k, -\sigma})
\ea
\\
with
\ba\nonumber
\gamma_{\vec k} = \frac{1}{2} (cos k_{x} + cosk_{y})
\ea
The third term in this expression is complex since $\mu = i \pi/ 2 \beta$. In this 
representation it is not possible to find a unitary Bogolioubov transformation which 
diagonalizes $H$. Hence a rigorous implementation of the constraint on the particle 
number per site is not possible.\\ 

Reasons for this situation may be the fact that the Hamiltonian contains terms with two 
particles with opposite spin created or annihilated on the same site which is incompatible 
with the fact that such configurations are not allowed in the present scheme. Terms 
of this type are typical in mean-field pairing Hamiltonians which lead to a non-conservation 
of the number of particles of the system.\\

\subsection{Lagrange multiplier approximation }

If the sites are occupied by one electron in the average the Lagrange procedure works.  
We do not develop the derivation of the mean-field here since it has been done elsewhere
~\cite{aue}.\\

\section{Conclusions.}

In summary we have shown that a strict constraint on the site occupation of a lattice quantum 
spin system described by Heisenberg-type models shows a sizable quantitative different
localization of the critical temperature when compared with the outcome of an average occupation 
constraint. Consequently it generates sizable effects on the behaviour of  order parameters 
and other physical observables.  This is true both in the case of antiferromagnetic (N\'eel)
and spin state symmetry.\\

Due to the complexity of quantum spin systems the choice of a physically meaningful mean field 
may depend on the coupling strengths of the model which describes the systems~\cite{wwz}. A $specific$ 
mean-field solution may even be a naive way to fix the "classical" contribution to the partition function 
which may in fact contain a mixture of different types of states. As already mentioned many efforts 
have been and are done in order to analyze and overcome these problems, see f.i.~\cite{tan,her,wen}. 
It is our aim to repeat the analysis we did above in the framework of a description which is able to 
implement more appropriate descriptions at the mean-field level.\\  

In a more realistic analysis one should of course take care of the contributions of quantum 
fluctuations which may be of overwhelming importance particularly in the vicinity of critical 
points. We expect to work out the first order contributions in a loop expansion for both types 
of constraints and symmetries considered in the present work in order to analyze and compare their 
relative importance as a function of temperature. Such an analysis may provide some view on the 
relative contribution of quantum corrections to order parameters and other observables for the 
different chosen mean-field ans\"atze.\\  

The authors would like to thank D. Cabra and T. Vekua for interesting and helpful discussions during 
the time of preparation of the present work.

\end{document}